\documentclass[sigconf,review=false]{acmart}
\usepackage{graphicx}
\usepackage{subcaption}
\usepackage{multirow}
\usepackage{float}
\usepackage{hyperref}

\copyrightyear{2025}
\acmYear{2025}
\setcopyright{rightsretained}
\acmConference[UMAP Adjunct '25]{Adjunct Proceedings of the 33rd ACM Conference on User Modeling, Adaptation and Personalization}{June 16--19, 2025}{New York City, NY, USA}
\acmBooktitle{Adjunct Proceedings of the 33rd ACM Conference on User Modeling, Adaptation and Personalization (UMAP Adjunct '25), June 16--19, 2025, New York City, NY, USA}
\acmDOI{10.1145/3708319.3733686}
\acmISBN{979-8-4007-1399-6/2025/06}

\begin{document}

\title[BIRD: A Museum Open Dataset Combining Behavior Patterns and Identity Types to Better Model Visitors' Exp.]{BIRD: A Museum Open Dataset Combining Behavior Patterns and Identity Types to Better Model Visitors' Experience}

\author{Alexanne Worm}
 \email{alexanne.worm@loria.fr}
 \affiliation{
   \institution{University of Lorraine (IDMC) - CNRS - LORIA}
   \city{Vandoeuvre-lès-Nancy}
   \country{FRANCE}
}
\author{Florian Marchal}
 \email{florian.marchal.bornert@gmail.com}
 \affiliation{
   \institution{University of Lorraine - CNRS - LORIA}
   \city{Vandoeuvre-lès-Nancy}
   \country{FRANCE}
}
\author{Sylvain Castagnos}
 \email{sylvain.castagnos@loria.fr}
 \affiliation{
   \institution{University of Lorraine - CNRS - LORIA}
   \city{Vandoeuvre-lès-Nancy}
   \country{FRANCE}
}
\date{May 2025}

\begin{abstract}
Lack of data is a recurring problem in Artificial Intelligence, as it is essential for training and validating models. This is particularly true in the field of cultural heritage, where the number of open datasets is relatively limited and where the data collected does not always allow for holistic modeling of visitors' experience due to the fact that data are {\it ad hoc} ({\it i.e.} restricted to the sole characteristics required for the evaluation of a specific model). To overcome this lack, we conducted a study between February and March 2019 aimed at obtaining comprehensive and detailed information about visitors, their visit experience and their feedback. We equipped 51 participants with eye-tracking glasses, leaving them free to explore the 3 floors of the museum for an average of 57 minutes, and to discover an exhibition of more than 400 artworks. On this basis, we built an open dataset combining contextual data (demographic data, preferences, visiting habits, motivations, social context\ldots), behavioral data (spatiotemporal trajectories, gaze data) and feedback (satisfaction, fatigue, liked artworks, verbatim\ldots). Our analysis made it possible to re-enact visitor identities combining the majority of characteristics found in the literature~\cite{sparacino_2002,hatala_2005,zancanaro_2007, falk_2016,kontiza_2018,grazioso_2020} and to reproduce the Veron and Levasseur profiles~\cite{veron_1989}. This dataset will ultimately make it possible to improve the quality of recommended paths in museums by personalizing the number of points of interest (POIs), the time spent at these different POIs, and the amount of information to be provided to each visitor based on their level of interest. Dataset URL: https://mbanv2.loria.fr/
\end{abstract}

\ccsdesc[500]{Human-centered computing~User studies}
\ccsdesc[500]{Human-centered computing~User models}
\ccsdesc[300]{Applied computing~Arts and humanities}
\ccsdesc[100]{Information systems~Clustering}
\ccsdesc[100]{Information systems~Location based services}

\keywords{Dataset, Identity-Related Data, Spatiotemporal Data, Gaze Data, Museum Visitors' Behavior, User Modeling, Recommenders}

\maketitle

\section{Introduction and Related Work}

Museums are often faced with the question of whether they should impose a narrative to visitors, for example through appropriate signage or audio tours, or whether they should leave them free to explore their environment. Recommender systems offer an interesting alternative between these two approaches, by proposing personalized routes both based on the individual expectations and the scenography of the museum~\cite{najbrt_2016}. To improve the performance of such systems, it is necessary to effectively exploit data of a very varied nature to model and understand visitors' behaviors in detail~\cite{ceccarelli_2024}. Collecting visitor data in the museum context is a widely studied subject, and many technologies are dedicated to this task~\cite{kovavisaruch_2012,seidenari_2017,ferrato_2022}. Nevertheless, there are few publicly available datasets that can be exploited for recommendation purposes.
Several types of data can be collected for in-depth analysis of visitor identities. Furka {\it et al.}~\cite{furka_2022} consider demographic characteristics and artwork ratings given by visitors in order to predict the future ones.
Packer and Roy~\cite{packer_2002} and Falk~\cite{falk_2016} are particularly interested in visitor identities, including their motivations and their learning experience in a museum. Data is collected via questionnaires.
Zancanaro {\it et al.}~\cite{zancanaro_2007} collected data on the visited artworks, including order, time spent and percentage observed. A particular focus is also made on the overall visitors' behavior based on their trajectories. 
Sparacino~\cite{sparacino_2002} manually retrieves visitor trajectories, annotating the visit with the number and duration of stops, as well as the items observed (12 in total). As an extension of this work, Hatala and Wakkary~\cite{hatala_2005} also collect interaction history (discrete time-space points of locations and selection of objects), user type according to Sparacino's nomenclature and user interests.
Yoshimura {\it et al.}~\cite{yoshimura_2012} propose an automated approach: bluetooth technology is used with sensors. The trajectories consist of a sequence of sensors that detected visitors. Girolami  {\it et al.}~\cite{girolami_2024} also use Bluetooth beacons to obtain approximate trajectories for a visit around 10 works with 32 visitors. Similar techniques are employed by Lanir  {\it et al.}~\cite{lanir_2017} to get global trajectories of visitors, with radio frequency based positioning system.
Another possibility to access and understand visitor behavior in an indoor environment consist to use data from another context, such as a conference~\cite{zhao_2021} or a shopping mall~\cite{brvsvcic_2013}. However, such domain transpositions do not capture the specific characteristics of a museum, in particular the cultural surroundings that are introduced and not found in other spaces.
At last, Grazioso {\it et al.}~\cite{grazioso_2020} investigate how mobile eye trackers allow a deeper understanding of visitors' behavior while they observe artworks. They collected gaze data from participants from two sessions, before and after a course in art history.

As we can observe in the literature, the analyzed datasets focus on only a few aspects of each visitor. Moreover, these datasets are rarely made public, which restricts their use. 
In this paper, we propose a dataset with a variety of visitor information to capture their complete profile. The data collected includes the complete trajectories of visitors, the works viewed and appreciated, and data on their identity (motivation, art knowledge...). The museum journeys of 51 visitors are made available in our dataset called BIRD so far, together with museum information (floor plans, artwork descriptions, daily statistics). BIRD is the acronym for Behavioral and Identity-Related Dataset. We plan to progressively increase the number of participants.

As far as we know, no dataset to exploit the different characteristics of visitors for recommendation and identity analysis in an indoor environment has been produced and made public.
The remainder of this paper is organized as follows. In Section~\ref{sec:procedure}, we provide the procedure of data collection and the pre-processing implemented to obtain a correctly formatted dataset. We then overview the dataset and their suitability for visitor identities in Section~\ref{sec:overview}. Finally, Section~\ref{sec:perspectives} is dedicated to the dataset benefits, conditions of use and future work.

\section{Experiment Procedure}\label{sec:procedure}
\subsection{Origin and landmarks}

The creation of this dataset is part of the MBANv2 project\footnote{https://mbanv2.loria.fr/} in collaboration with the Nancy Museum of Fine Arts\footnote{https://musee-des-beaux-arts.nancy.fr/en/museum}, which boasts a varied collection of artworks, and welcomes around 300,000 visitors every year. This project has two major objectives: firstly, to create a dataset that is as complete and faithful as possible to capture the visitors' experience, and secondly, to create a simulation and test environment (reproduction of a real museum in Unity). These two objectives enable researchers to develop and compare different machine learning models: crowd and trajectory simulation, recommender systems and virtual guides. User studies could also be carried out in the virtual environment to gather information about them and analyze their behavior.
This article is dedicated to the first objective of the project.

\subsection{Data acquisition}

Our study involved 51 visitors and more than 400 pieces of artwork from 16th to 21th centuries. Participants were free to explore the 3 floors of the Nancy Museum of Fine Arts as they wished, without any influence such as a guide or a recommender system. 
Each visitor was approached at the beginning of the visit with the agreement of the museum. After being informed on the purpose of the study and data collected, they were asked to complete a consent form and were invited to visit the museum as they would under normal conditions. The experimental protocol was declared and authorized by the Data Protection Officer of our university to guarantee its legal compliance.

The museum contains a large variety of artworks (paintings, sculptures\ldots). The scenography is thought out chronologically, with each floor dedicated to a specific period, and each room to a specific art movement (impressionism, post-impressionism, cubism\ldots). The spatial topology also differs from one floor to another. The first floor gives more leeway to visitors through big rooms and multiple routes to reach the same painting. The second and third floors constrain the path with rooms only accessible by few entries. 
The layout of the museum and the position of the artworks were mapped out to determine visitor trajectories. The map is available in PDF and JSON formats.
The dataset of artworks has also been created, with information on each item (period, theme, description, etc.). Information was collected via museum panels, the Nancy Museum of Fine Arts database, and specialized websites.

\begin{figure}[]
\centering
	\includegraphics[width=1\linewidth] {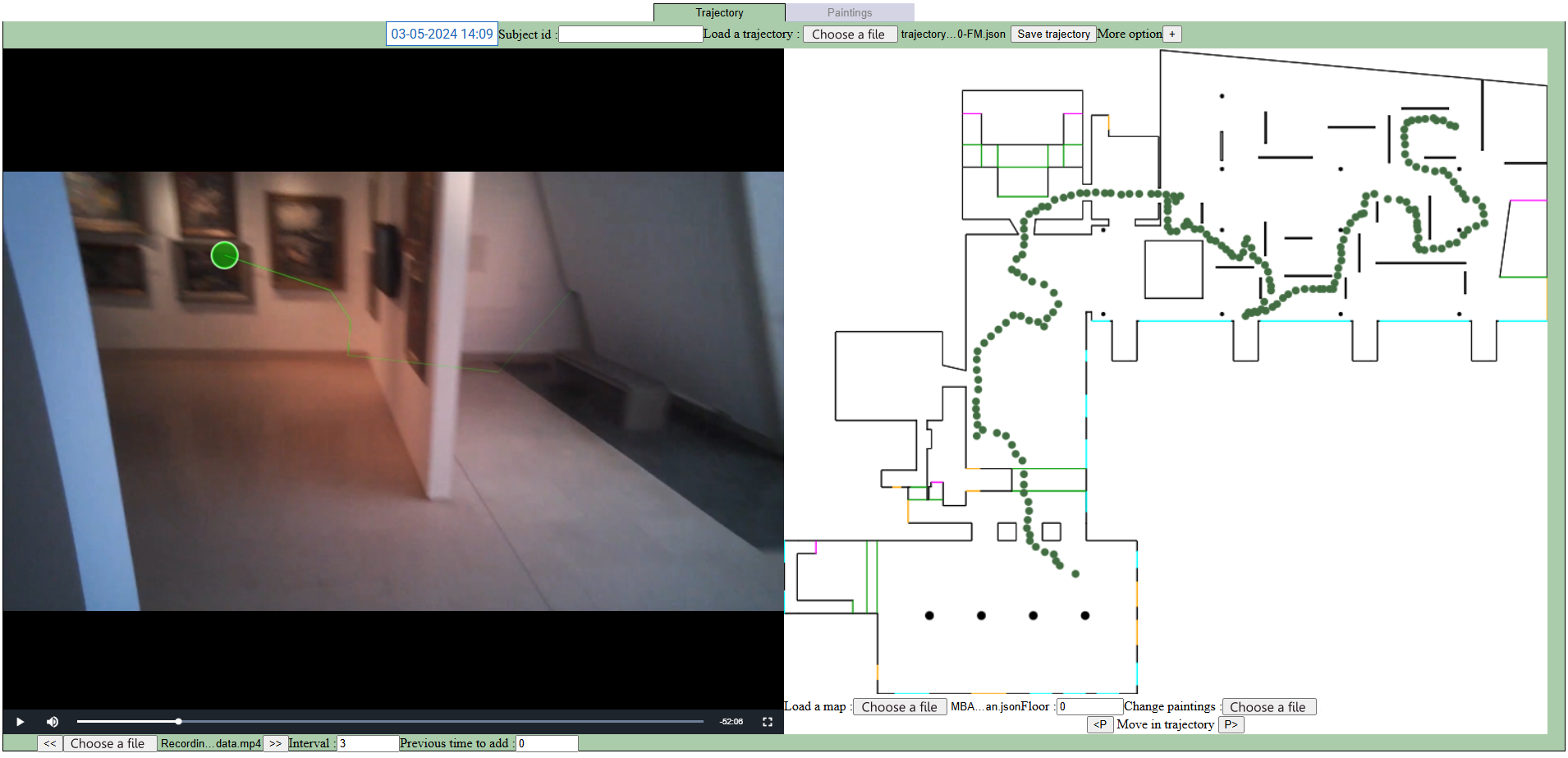}  
	\caption{Illustration of the platform used to obtain trajectories.}
	\label{fig:kiwimapper}
        \Description{Screenshot showing an example of a trajectory creation by using the platform elaborated by the team.}
\end{figure}

Tobii Glasses 2 (100Hz) were employed to anonymously acknowledge the items observed and the path taken by the visitors, respectively from its eye-tracking cameras and its scene camera. We developed a web platform to get the trajectories and list of items from the videos, by manually pointing the visitor's position on the map in JSON format (NMFA\_3floors\_plan.json). At the end of the process, a JSON file named after his anonymous unique ID is generated for each visitor. An illustration of the process is visible in Figure~\ref{fig:kiwimapper}. Another window is dedicated to the acquisition of items. In this situation, all the walls containing items are clickable. When a visitor notices an item, it can be added to the list by clicking on the wall and select the correct image.
Only paintings have been included in the list of visible artworks. It is possible that the visitor stops at a place to observe an item of another kind (sculpture\ldots) but these items have not been taken into account in the analysis.
The lists of artworks observed were divided into two groups, in order to obtain the duration of viewing an artwork: one group corresponds to the beginning of viewing a work, while the other corresponds to the end.
To make it easier to use the trajectories and list of items with multiple languages, the JSON files obtained from the platform have been converted into CSV files (items\_idVisitor.csv, items\_idVisitor\_end.csv, trajectory\_idVisitor.csv).

As well as collecting trajectories and list of items, pre- and post-questionnaires were drawn up to supplement the information we were able to obtain about visitors. They were completed at the beginning and end of the visit. Visitors could choose whether or not to complete the questionnaires. All decided to fill in both documents. The questionnaires were designed to obtain data on different aspects of the visitor's identity: demographic (age, gender, diploma\ldots), physiological (physical fatigue\ldots), psychological (motivations, reasons to come, crowd tolerance\ldots) and group factor (accompanied or not). Details of information acquired from the questionnaires are shown in the subsection Format and Key features (see Section~\ref{subsec:key_features}). 

At the end of the visit, after filling in the questionnaires, visitors were invited to use a platform to select items they liked on a mosaic of all the photographs of the artworks in the museum, randomly distributed by floor. Visitors were free to choose the artworks they liked, regardless of whether they had seen them during the visit or not. This step allows us to elicit the artistic preferences of visitors, but also to measure the coverage rate of their visit relative to their preferences (items likely to interest them and which they did not notice during their exploration). Each list of items selected by a visitor was then retrieved and placed in a CSV file (explicit\_feedback\_visitors.csv).

\subsection{Data pre-processing}

\begin{figure}[]
    \centering
    \begin{subfigure}{0.2\textwidth}
        \centering
        \includegraphics[width=\linewidth]{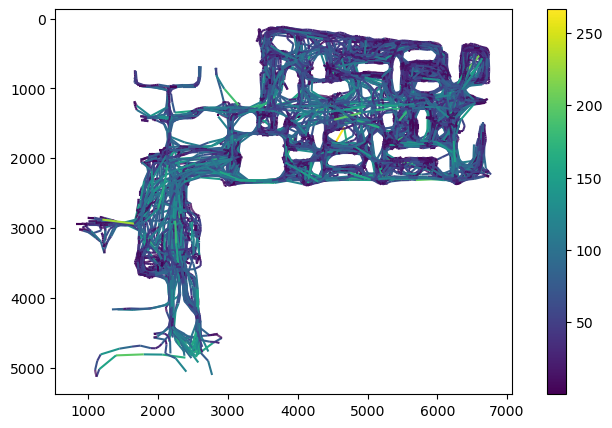}
        \caption{Normalized trajectories of the first floor.}
        \label{fig:groundNorm}
    \end{subfigure}
    \hfill 
    \begin{subfigure}{0.2\textwidth}
        \centering
        \includegraphics[width=\linewidth]{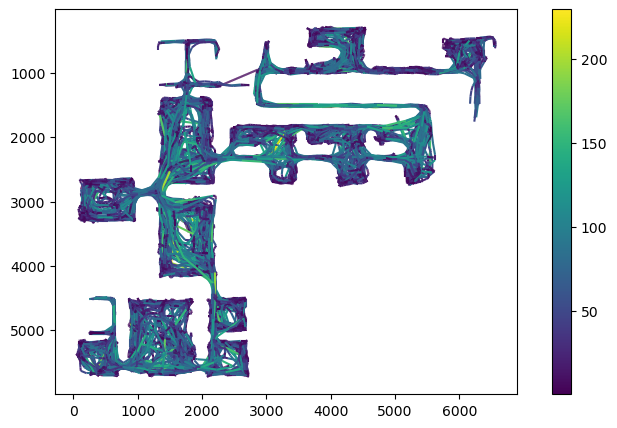}
        \caption{Normalized trajectories of the second floor.}
        \label{fig:firstNorm}
    \end{subfigure}
    \hfill
    \begin{subfigure}{0.2\textwidth}
        \centering
        \includegraphics[width=\linewidth]{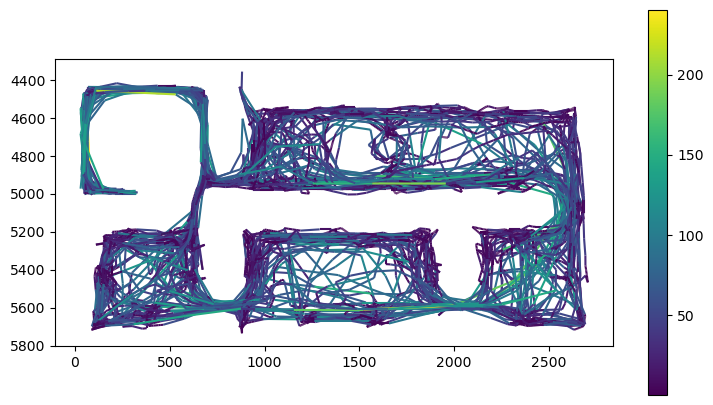}
        \caption{Normalized trajectories of the third floor.}
        \label{fig:secondNorm}
    \end{subfigure}
    \caption{Normalized trajectories. The color corresponds to speed, in pixel/sec (unit from the platform). Axes also have pixel units and each point corresponds to a visitor's position at a specific timestamp. All trajectories are represented in these figures. 100 pixels correspond to 1 meter.}
    \label{fig:normTrajs}
    \Description{Illustration of the trajectories normalized for each floor in the museum. The color corresponds to speed, in pixel/sec (unit from the platform) from blue (low speed) to yellow (high speed). Axes also have pixel units. 100 pixels correspond to 1 meter. }
\end{figure}

Raw trajectories may contain noise, as they are produced manually and can sometimes be too precise, preventing the detection of global patterns. To remove this noise, we normalize the trajectories.
We relied on the MovingPandas library v0.20.0~\cite{graser_movingpandas_2019} to normalize the trajectories with an interval of two seconds between two different positions.
We obtain consistent trajectories for each floor as observed in \ref{fig:normTrajs}.

Each cleaned trajectory is available along the raw trajectories in the dataset. An important added value of our dataset compared to the state-of-the-art is that we know precisely the direction and speed of movement of visitors in real time, whereas the datasets mentioned in Introduction only have the positions of participants at different timestamps.

From each trajectory, we can extract a set of features to further analyze the visitor's behavior by relying on MovingPandas. In our study, we chose to select a few characteristics to check its suitability for some museum identities, such as those of Veron and Levasseur ~\cite{veron_1989}. 
We therefore collected the duration of each trajectory, the average speed, the number of stops, the length of the trajectory and the number of items observed. This processing is also available in a CSV file (semantic\_info\_entire\_trajectories.csv).

\subsection{Format and key features}\label{subsec:key_features}

Based on the data collected, we have created a dataset with various files providing access to specific visitor information. 

The trajectories of the N visitors are presented in a sequence of tuples containing for each timestamp the position of the visitor $v$:
\begin{math}
    T_v = \{(t, fl, x, y) | t = 0,..., t_{end}; fl \in (0,1,2); (x,y) \in R^2 \}, \forall v \in [1,..., N]
\end{math}, $t_{end}$ the timestamp of the last point of the trajectory. The coordinates ($x$ and $y$) of the visitor for each floor are in pixels, corresponding to the web platform unit (100 pixels = 1 meter). $fl$ corresponds to the floor.

The list of items seen (items\_idVisitor.csv, items\_idVisitor\_end.csv) are also represented by a sequence of tuples for each visitor $v$: 
\begin{math}
    I_v = \{(t, fl, paintingID) | t = 0,..., t_{end}; fl \in (0,1,2)\}
\end{math}
, $paintingID$ a string corresponding to the ID of the artwork image.
It should be noted that lists with the beginning of items seen may contain more items than end lists. These items which are not present in the end list, have been seen for a too short duration to be included in the end list.

The dataset giving the global characteristics of the trajectories is composed as in Table~\ref{tab:dataset_traj_info}.

\begin{table}[]
  \caption{Data description of the dataset with information on trajectories.}
  \Description{Description of the variables used to describe the trajectories and their type. There are 6 variables: speed (in pixels/sec), duration (in seconds), number of items (float), number of stops (integer), length (in pixels) and trajectory\_id.  }
  \label{tab:dataset_traj_info}
  \begin{tabular}{ccl}
    \toprule
    Name&Type\\
    \midrule
    Trajectory id & Integer\\
    Duration & Float (in seconds)\\
    Speed & pixels/sec (100 pixels = 1 meter)\\
    Nb\_items & Float\\
    Nb\_stops & Integer\\
    Length & Float (in pixels, 100 pixels = 1 meter)\\
  \bottomrule
\end{tabular}
\end{table}

Concerning the questionnaires, 23 questions were asked in the pre-questionnaire, which was available in French and in English. The focus is on demographic data (age, qualifications, gender), visit preferences, visit frequency, motivation, objectives, physiological and psychological data (fatigue, distance tolerance, crowd tolerance, etc.). The items related to the objectives were defined on the basis of Falk's identities (\cite{falk_2016}). 
For the post-questionnaire, also in French and English, 29 questions were elaborated. This time, the main topics were art expertise, the benefits of new technologies (applications, recommender systems), the physical and psychological states at the end of the visit, the information about artworks memorized at the end of the visit, and other comments on the visitor's experience. 

Finally, the artworks appreciated by visitors correspond of a file of tuples of this form: \{visitor\_id, id\_item, timeOfSelection\}.

The project presentation and access to the dataset are available at the following address:  
\url{https://mbanv2.loria.fr/}

\section{Dataset overview}\label{sec:overview}
\subsection{Statistics}
\begin{figure}
    \centering
    \begin{subfigure}{0.23\textwidth}
        \centering
        \includegraphics[width=\linewidth]{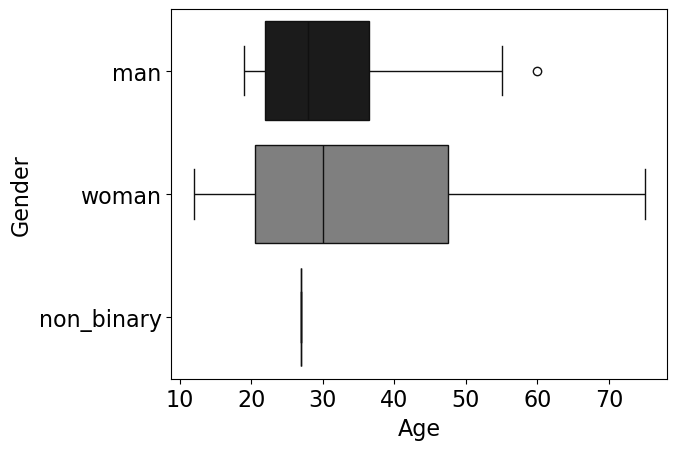}
        \caption{Age distribution of visitors.}
    \end{subfigure}
    \hfill 
    \begin{subfigure}{0.22\textwidth}
        \centering
        \includegraphics[width=\linewidth]{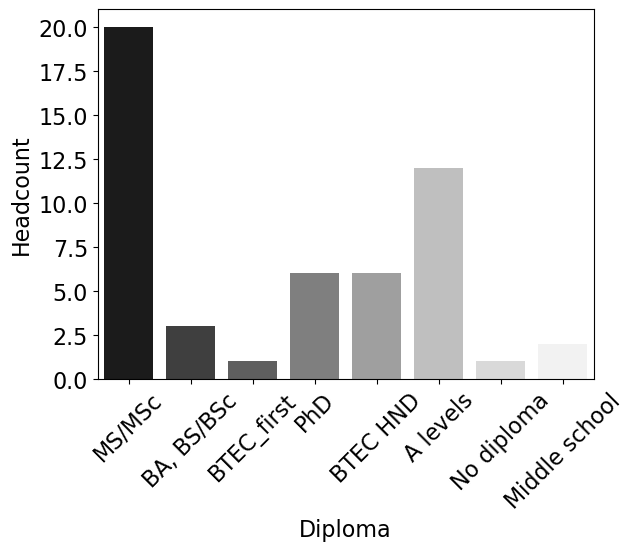}
        \caption{Educational attainment  distribution of visitors.}
    \end{subfigure}
    \caption{Statistics of the visitor identities.}
    \label{fig:visitorProfiles}
    \Description{Figures describing the statistics of visitor identities of the dataset.} 
\end{figure}

The dataset shows a diversity of visitor identities. 51 visitor identities are currently available. 
Visitor statistics can be seen in Figure~\ref{fig:visitorProfiles}. 
The experiment included 27 males, 23 females and 1 non-binary person. The mean age is 33 years, going from 12 to 75 years old.

The statistical study provides some information on the 51 developed trajectories. The average visit duration is 57.6 minutes, with a minimum duration of 11 minutes and a maximum duration of 1 hour and 55 minutes. 144 items per visitor are seen on average, with a speed of around 0.26 m/s. This speed is low in view of the environment, which causes many stops (approximately 54 stops per visit). The average length of a trajectory is 838 meters, with a minimum of 254 meters and a maximum of 1361 meters.
Each visitor spends an average of 29 seconds in front of selected artworks.

\subsection{Suitability for museum identity analysis}

\begin{figure}[]
\centering
	\includegraphics[width=1\linewidth]{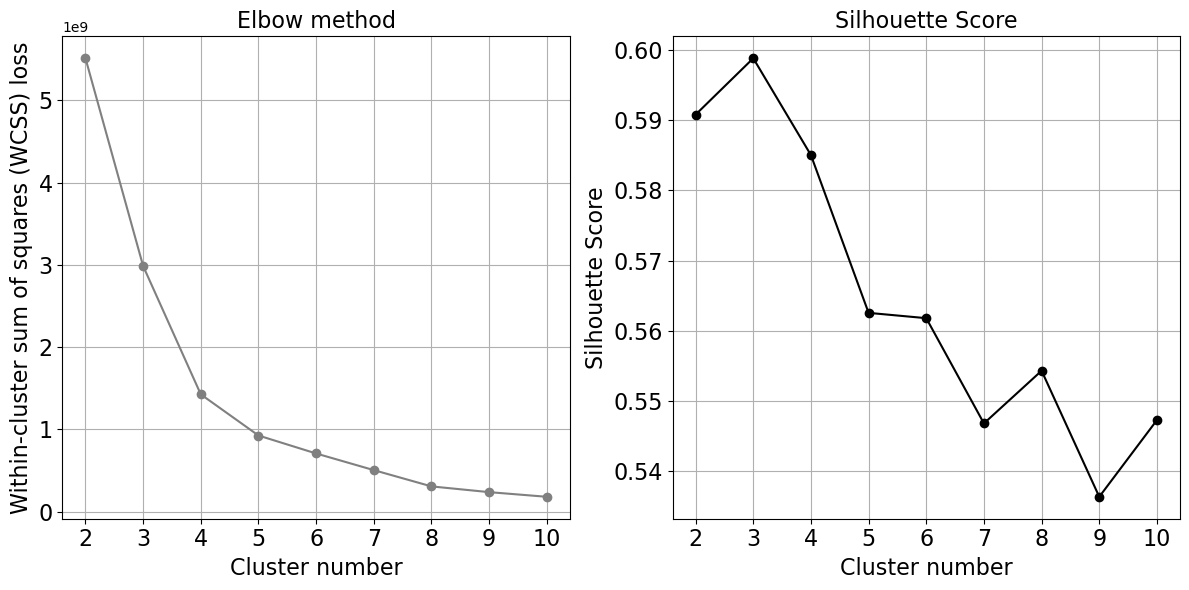}  
	\caption{Results obtained from Kmeans clustering on the trajectories information (length, number of items seen, number of stops, speed and duration).}
	\label{fig:clustering}
        \Description{Values of Silhouette Score and WCSS considering different number of clusters for Kmeans Clustering on trajectory information. As observed in the figure and considering the two employed indexes, 4 clusters seem to be relevant. }
\end{figure}

To verify the suitability of our dataset for the study of visitor identities and as a proof of concept, we performed clustering to detect Veron and Levasseur profiles~\cite{veron_1989}. To do this, we used data from the file containing global information on trajectories (semantic\_info\_entire\_trajectories). The clustering technique employed is Kmeans. We used the WCSS loss with the elbow method and Silhouette score, as well as the Davies-Bouldin Index and Calinski-Harabasz Index to select the most relevant number of clusters. As shown in Figure~\ref{fig:clustering}, and considering the index values, 4 seems to be the most appropriate number of clusters. A closer look at the centroids obtained reveals the nature of each group: Grasshopper, Ant, Fish and Butterfly.

An interesting future work would be to add information on the order of visit (chronological or not), in order to check whether this could have an impact on the classification.

\section{License, Benefits and Perspectives}\label{sec:perspectives}

This dataset provides access to detailed visitor identities with information that can be used in combination or separately. It can be employed for a wide range of applications, including the study of visitor identities and behaviors during visits, trajectory prediction, crowd simulation, use in Natural Language Processing (study of visitor comments...). Recommender systems are particularly well-suited to this type of data. Several types of systems can be used (Content-Based Filtering, Collaborative Filtering, Trajectory recommendation...) and compared thanks to this BIRD dataset. In other words, this dataset has been built independently of any research hypothesis or model to be evaluated, in order to be as generic as possible and useful to the research community, as MovieLens used to be in another application domain.
Nevertheless, it is important to note that these data were collected in a specific context and over a limited period of time. The dataset seems to be in line with a certain part of the literature (Veron and Levasseur profiles \cite{veron_1989}), and one of our future works will reside in verifying whether this dataset can be generalized to other situations. 

Other future works will consist in enriching the dataset with more visitors, while giving access to gaze data.
Our dataset will also be enriched with more behavioral and identity-related data (isovists, artwork complexity...) so as to train a sequence-based recommender system and test its capacity to predict visiting styles over time. We believe that this information will enable the system to tailor recommendations according to the visible items and the cognitive load that the artworks can bring.
To enable this dataset to be used in Deep Learning systems, a data augmentation technique will be used to enlarge it while trying to maintain the coherence of each visitor, in particular their behaviors and trajectories.
The complete dataset will be made available, along with the code used to obtain trajectories, information and analysis results under CC-BY-NC-SA 4.0 license. Any use of this dataset for research purposes must be accompanied by a citation of this paper.

\begin{acks}
This research was supported by the non-economic valuation project called MBANv2. It was the subject of an agreement signed by the University of Lorraine, the metropolitan area of Nancy (Ville de Nancy) and the Nancy Museum of Fine Arts. We would like to thank Sophie Mouton, Sophie Toulouze, Charles Villeneuve de Janti, Michèle Leinen, and Jean-Paul Darada for providing information on the artworks and for authorizing this study to be conducted within the museum.
\end{acks}

\bibliographystyle{ACM-Reference-Format}
\bibliography{bibfile}


\begin{thebibliography}{20}


\ifx \showCODEN    \undefined \def \showCODEN     #1{\unskip}     \fi
\ifx \showDOI      \undefined \def \showDOI       #1{#1}\fi
\ifx \showISBNx    \undefined \def \showISBNx     #1{\unskip}     \fi
\ifx \showISBNxiii \undefined \def \showISBNxiii  #1{\unskip}     \fi
\ifx \showISSN     \undefined \def \showISSN      #1{\unskip}     \fi
\ifx \showLCCN     \undefined \def \showLCCN      #1{\unskip}     \fi
\ifx \shownote     \undefined \def \shownote      #1{#1}          \fi
\ifx \showarticletitle \undefined \def \showarticletitle #1{#1}   \fi
\ifx \showURL      \undefined \def \showURL       {\relax}        \fi
\providecommand\bibfield[2]{#2}
\providecommand\bibinfo[2]{#2}
\providecommand\natexlab[1]{#1}
\providecommand\showeprint[2][]{arXiv:#2}

\bibitem[Br{\v{s}}{\v{c}}i{\'c} et~al\mbox{.}(2013)]%
        {brvsvcic_2013}
\bibfield{author}{\bibinfo{person}{Dra{\v{z}}en Br{\v{s}}{\v{c}}i{\'c}},
  \bibinfo{person}{Takayuki Kanda}, \bibinfo{person}{Tetsushi Ikeda}, {and}
  \bibinfo{person}{Takahiro Miyashita}.} \bibinfo{year}{2013}\natexlab{}.
\newblock \showarticletitle{Person tracking in large public spaces using 3-D
  range sensors}.
\newblock \bibinfo{journal}{\emph{IEEE Transactions on Human-Machine Systems}}
  \bibinfo{volume}{43}, \bibinfo{number}{6} (\bibinfo{year}{2013}),
  \bibinfo{pages}{522--534}.
\newblock


\bibitem[Ceccarelli et~al\mbox{.}(2024)]%
        {ceccarelli_2024}
\bibfield{author}{\bibinfo{person}{Sofia Ceccarelli}, \bibinfo{person}{Amedeo
  Cesta}, \bibinfo{person}{Gabriella Cortellessa}, \bibinfo{person}{Riccardo
  De~Benedictis}, \bibinfo{person}{Francesca Fracasso}, \bibinfo{person}{Laura
  Leopardi}, \bibinfo{person}{Luca Ligios}, \bibinfo{person}{Ernesto Lombardi},
  \bibinfo{person}{Malatesta~Saverio Giulio}, \bibinfo{person}{Angelo Oddi},
  \bibinfo{person}{Alfonsina Pagano}, \bibinfo{person}{Augusto Palombini},
  \bibinfo{person}{Gianmauro Romagna}, \bibinfo{person}{Marta Sanzari}, {and}
  \bibinfo{person}{Marco Schaerf}.} \bibinfo{year}{2024}\natexlab{}.
\newblock \showarticletitle{Evaluating visitors' experience in museum:
  comparing artificial intelligence and multi-partitioned analysis}.
\newblock \bibinfo{journal}{\emph{Digital Applications in Archaeology and
  Cultural Heritage}}  \bibinfo{volume}{33} (\bibinfo{year}{2024}).
\newblock
\urldef\tempurl%
\url{https://doi.org/10.1016/j.daach.2024.e00340}
\showDOI{\tempurl}


\bibitem[Falk(2016)]%
        {falk_2016}
\bibfield{author}{\bibinfo{person}{John~H. Falk}.}
  \bibinfo{year}{2016}\natexlab{}.
\newblock \bibinfo{booktitle}{\emph{Identity and the museum visitor
  experience}}.
\newblock \bibinfo{publisher}{Routledge}.
\newblock


\bibitem[Ferrato et~al\mbox{.}(2022)]%
        {ferrato_2022}
\bibfield{author}{\bibinfo{person}{Alessio Ferrato}, \bibinfo{person}{Carla
  Limongelli}, \bibinfo{person}{Mauro Mezzini}, {and} \bibinfo{person}{Giuseppe
  Sansonetti}.} \bibinfo{year}{2022}\natexlab{}.
\newblock \showarticletitle{Using deep learning for collecting data about
  museum visitor behavior}.
\newblock \bibinfo{journal}{\emph{Applied Sciences}} \bibinfo{volume}{12},
  \bibinfo{number}{2} (\bibinfo{year}{2022}), \bibinfo{pages}{533}.
\newblock


\bibitem[Furka(2022)]%
        {furka_2022}
\bibfield{author}{\bibinfo{person}{Bc~Marek Furka}.}
  \bibinfo{year}{2022}\natexlab{}.
\newblock \emph{\bibinfo{title}{Exhibit rating prediction and vistior path
  prediction in a museum setting}}.
\newblock \bibinfo{thesistype}{Ph.\,D. Dissertation}. \bibinfo{school}{Wien}.
\newblock


\bibitem[Girolami et~al\mbox{.}(2024)]%
        {girolami_2024}
\bibfield{author}{\bibinfo{person}{Michele Girolami}, \bibinfo{person}{Davide
  La~Rosa}, {and} \bibinfo{person}{Paolo Barsocchi}.}
  \bibinfo{year}{2024}\natexlab{}.
\newblock \showarticletitle{Bluetooth dataset for proximity detection in indoor
  environments collected with smartphones}.
\newblock \bibinfo{journal}{\emph{Data in Brief}}  \bibinfo{volume}{53}
  (\bibinfo{year}{2024}), \bibinfo{pages}{110215}.
\newblock


\bibitem[Graser(2019)]%
        {graser_movingpandas_2019}
\bibfield{author}{\bibinfo{person}{Anita Graser}.}
  \bibinfo{year}{2019}\natexlab{}.
\newblock \showarticletitle{{MovingPandas}: {Efficient} {Structures} for
  {Movement} {Data} in {Python}}.
\newblock \bibinfo{journal}{\emph{GI\_Forum ‒ Journal of Geographic
  Information Science}} \bibinfo{volume}{7}, \bibinfo{number}{1}
  (\bibinfo{year}{2019}), \bibinfo{pages}{54--68}.
\newblock
\showISSN{2308-1708, 2308-1708}
\urldef\tempurl%
\url{https://doi.org/10.1553/giscience2019_01_s54}
\showDOI{\tempurl}


\bibitem[Grazioso et~al\mbox{.}(2020)]%
        {grazioso_2020}
\bibfield{author}{\bibinfo{person}{Marco Grazioso}, \bibinfo{person}{Roberto
  Esposito}, \bibinfo{person}{Emma Maayan-Fanar}, \bibinfo{person}{Tsvi
  Kuflik}, {and} \bibinfo{person}{Francesco Cutugno}.}
  \bibinfo{year}{2020}\natexlab{}.
\newblock \showarticletitle{Using Eye Tracking Data to Understand Visitors'
  Behaviour}. In \bibinfo{booktitle}{\emph{AVI2CH Workshop on Advanced Visual
  Interfaces and Interactions in Cultural Heritage}}. \bibinfo{address}{Island
  of Ischia, Italy}.
\newblock


\bibitem[Hatala and Wakkary(2005)]%
        {hatala_2005}
\bibfield{author}{\bibinfo{person}{Marek Hatala} {and} \bibinfo{person}{Ron
  Wakkary}.} \bibinfo{year}{2005}\natexlab{}.
\newblock \showarticletitle{Ontology-Based User Modeling in an Augmented Audio
  Reality System for Museums}.
\newblock \bibinfo{journal}{\emph{User Modeling and User-Adapted Interaction}}
  \bibinfo{volume}{15}, \bibinfo{number}{3-4} (\bibinfo{year}{2005}),
  \bibinfo{pages}{339--380}.
\newblock


\bibitem[Kontiza et~al\mbox{.}(2018)]%
        {kontiza_2018}
\bibfield{author}{\bibinfo{person}{K. Kontiza}, \bibinfo{person}{O. Loboda},
  \bibinfo{person}{L. Deladiennee}, \bibinfo{person}{S. Castagnos}, {and}
  \bibinfo{person}{Y. Naudet}.} \bibinfo{year}{2018}\natexlab{}.
\newblock \showarticletitle{A Museum App to Trigger Users' Reflection}. In
  \bibinfo{booktitle}{\emph{2nd International Workshop on Mobile Access to
  Cultural Heritage (MobileCH)}}.
\newblock


\bibitem[Kovavisaruch et~al\mbox{.}(2012)]%
        {kovavisaruch_2012}
\bibfield{author}{\bibinfo{person}{La-or Kovavisaruch}, \bibinfo{person}{Virach
  Sornleardlumvanich}, \bibinfo{person}{Thatsanee Chalernporn},
  \bibinfo{person}{Pobsit Kamolvej}, {and} \bibinfo{person}{Nitirat
  Iamrahong}.} \bibinfo{year}{2012}\natexlab{}.
\newblock \showarticletitle{Evaluating and collecting museum visitor behavior
  via RFID}. In \bibinfo{booktitle}{\emph{2012 Proceedings of PICMET'12:
  Technology Management for Emerging Technologies}}. IEEE,
  \bibinfo{pages}{1099--1101}.
\newblock


\bibitem[Lanir et~al\mbox{.}(2017)]%
        {lanir_2017}
\bibfield{author}{\bibinfo{person}{Joel Lanir}, \bibinfo{person}{Tsvi Kuflik},
  \bibinfo{person}{Julia Sheidin}, \bibinfo{person}{Nisan Yavin},
  \bibinfo{person}{Kate Leiderman}, {and} \bibinfo{person}{Michael Segal}.}
  \bibinfo{year}{2017}\natexlab{}.
\newblock \showarticletitle{Visualizing museum visitors’ behavior: Where do
  they go and what do they do there?}
\newblock \bibinfo{journal}{\emph{Personal and Ubiquitous Computing}}
  \bibinfo{volume}{21} (\bibinfo{year}{2017}), \bibinfo{pages}{313--326}.
\newblock


\bibitem[Najbrt and Kapounová(2016)]%
        {najbrt_2016}
\bibfield{author}{\bibinfo{person}{Lukas Najbrt} {and} \bibinfo{person}{Jana
  Kapounová}.} \bibinfo{year}{2016}\natexlab{}.
\newblock \showarticletitle{Categorization of Museum Visitors as Part of System
  for Personalized Museum Tour}.
\newblock \bibinfo{journal}{\emph{International Journal of Information and
  Communication Technologies in Education}} \bibinfo{volume}{3},
  \bibinfo{number}{1} (\bibinfo{year}{2016}), \bibinfo{pages}{17--27}.
\newblock


\bibitem[Packer and Ballantyne(2002)]%
        {packer_2002}
\bibfield{author}{\bibinfo{person}{Jan Packer} {and} \bibinfo{person}{Roy
  Ballantyne}.} \bibinfo{year}{2002}\natexlab{}.
\newblock \showarticletitle{Motivational factors and the visitor experience: A
  comparison of three sites}.
\newblock \bibinfo{journal}{\emph{Curator: The Museum Journal}}
  \bibinfo{volume}{45}, \bibinfo{number}{3} (\bibinfo{year}{2002}),
  \bibinfo{pages}{183--198}.
\newblock


\bibitem[Seidenari et~al\mbox{.}(2017)]%
        {seidenari_2017}
\bibfield{author}{\bibinfo{person}{Lorenzo Seidenari}, \bibinfo{person}{Claudio
  Baecchi}, \bibinfo{person}{Tiberio Uricchio}, \bibinfo{person}{Andrea
  Ferracani}, \bibinfo{person}{Marco Bertini}, {and}
  \bibinfo{person}{Alberto~Del Bimbo}.} \bibinfo{year}{2017}\natexlab{}.
\newblock \showarticletitle{Deep artwork detection and retrieval for automatic
  context-aware audio guides}.
\newblock \bibinfo{journal}{\emph{ACM Transactions on Multimedia Computing,
  Communications, and Applications (TOMM)}} \bibinfo{volume}{13},
  \bibinfo{number}{3s} (\bibinfo{year}{2017}), \bibinfo{pages}{1--21}.
\newblock


\bibitem[Sparacino(2002)]%
        {sparacino_2002}
\bibfield{author}{\bibinfo{person}{Flavia Sparacino}.}
  \bibinfo{year}{2002}\natexlab{}.
\newblock \showarticletitle{The Museum Wearable: Real-Time Sensor-Driven
  Understanding of Visitors' Interests for Personalized Visually-Augmented
  Museum Experiences}. In \bibinfo{booktitle}{\emph{Museums and the Web}}.
  \bibinfo{address}{Boston, MA}.
\newblock


\bibitem[Ver{\'o}n and Levasseur(1989)]%
        {veron_1989}
\bibfield{author}{\bibinfo{person}{Eliseo Ver{\'o}n} {and}
  \bibinfo{person}{Martine Levasseur}.} \bibinfo{year}{1989}\natexlab{}.
\newblock \bibinfo{booktitle}{\emph{Ethnographie de l'exposition: l'espace, le
  corps et le sens}}.
\newblock \bibinfo{publisher}{Centre Georges Pompidou, Biblioth{\`e}que
  publique d'information}.
\newblock


\bibitem[Yoshimura et~al\mbox{.}(2012)]%
        {yoshimura_2012}
\bibfield{author}{\bibinfo{person}{Yuji Yoshimura}, \bibinfo{person}{Fabien
  Girardin}, \bibinfo{person}{Juan~Pablo Carrascal}, \bibinfo{person}{Carlo
  Ratti}, {and} \bibinfo{person}{Josep Blat}.} \bibinfo{year}{2012}\natexlab{}.
\newblock \showarticletitle{New tools for studying visitor behaviours in
  museums: a case study at the Louvre}.
\newblock In \bibinfo{booktitle}{\emph{Information and Communication
  Technologies in Tourism 2012}}. \bibinfo{publisher}{Springer},
  \bibinfo{pages}{391--402}.
\newblock


\bibitem[Zancanaro et~al\mbox{.}(2007)]%
        {zancanaro_2007}
\bibfield{author}{\bibinfo{person}{Massimo Zancanaro}, \bibinfo{person}{Tsvi
  Kuflik}, \bibinfo{person}{Zvi Boger}, \bibinfo{person}{Dina Goren-Bar}, {and}
  \bibinfo{person}{Dan Goldwasser}.} \bibinfo{year}{2007}\natexlab{}.
\newblock \showarticletitle{Analyzing Museum Visitors' Behavior Patterns}. In
  \bibinfo{booktitle}{\emph{User Modeling}}. \bibinfo{publisher}{Springer
  Berlin Heidelberg}, \bibinfo{address}{Berlin, Heidelberg},
  \bibinfo{pages}{238--246}.
\newblock


\bibitem[Zhao et~al\mbox{.}(2021)]%
        {zhao_2021}
\bibfield{author}{\bibinfo{person}{Ying Zhao}, \bibinfo{person}{Xin Zhao},
  \bibinfo{person}{Siming Chen}, \bibinfo{person}{Zhuo Zhang}, {and}
  \bibinfo{person}{Xin Huang}.} \bibinfo{year}{2021}\natexlab{}.
\newblock \showarticletitle{An indoor crowd movement trajectory benchmark
  dataset}.
\newblock \bibinfo{journal}{\emph{IEEE Transactions on Reliability}}
  \bibinfo{volume}{70}, \bibinfo{number}{4} (\bibinfo{year}{2021}),
  \bibinfo{pages}{1368--1380}.
\newblock


\end{thebibliography}

\end{document}